\documentclass[a4paper,11pt]{article}
\pdfoutput=1

\usepackage{jheppub} 
\usepackage{natbib}
\setcitestyle{square,comma,numbers,sort&compress}
\usepackage{enumerate}
\usepackage{hyperref}
\usepackage{tabularx,booktabs}
\usepackage[dvipsnames]{xcolor}
\usepackage{comment}
\usepackage[caption=false]{subfig}
\usepackage{cancel}
\usepackage{fontawesome}
\usepackage{hyperref}

\usepackage[section]{placeins}

\newcolumntype{Y}{>{\centering\arraybackslash}X}

\definecolor{myRED}{rgb}{0.8, 0.25, 0.33}

\newcommand{\rmv}[1]{{\iffalse #1 \fi}}

\usepackage{mathtools}

\usepackage[force]{feynmp-auto}
\usepackage[T1]{fontenc} 

\usepackage{cleveref}

\title{Type-I two-Higgs-doublet model and gravitational waves from domain walls bounded by strings}

\author[a]{Bowen Fu,}
\author[b]{Anish Ghoshal,}
\author[c]{Stephen F. King,} 
\author[d]{and Moinul Hossain Rahat} 
\affiliation[a]{Tsung-Dao Lee Institute, Shanghai Jiao Tong University, Shanghai 200240, China}
\affiliation[b]{Institute of Theoretical Physics, Faculty of Physics, University of Warsaw, ul. Pasteura 5, 02-093
Warsaw, Poland}
\affiliation[c]{School of Physics \& Astronomy, University of Southampton, Southampton SO17 1BJ, UK}
\affiliation[d]{Instituto de F\'isica Corpuscular, Universidad de Valencia and CSIC, Edificio Institutos Investigaci\'on, C/Catedr\'atico
Jos\'e Beltr\'an 2, 46980 Paterna, Spain}

\emailAdd{bowen\_fu@sjtu.edu.cn}
\emailAdd{anish.ghoshal@fuw.edu.pl}
\emailAdd{king@soton.ac.uk}
\emailAdd{moinul.rahat@ific.uv.es}

\abstract{
The spontaneous breaking of a $U(1)$ symmetry via an intermediate discrete symmetry may yield a hybrid topological defect of \emph{domain walls bounded by cosmic strings}. The decay of this defect network leads to a unique gravitational wave signal spanning many orders in observable frequencies, that can be distinguished from signals generated by other sources. We investigate the production of gravitational waves from this mechanism in the context of the type-I two-Higgs-doublet model extended by a $U(1)_R$ symmetry, that simultaneously accommodates the seesaw mechanism, anomaly cancellation, and eliminates flavour-changing neutral currents. The gravitational wave spectrum produced by the string-bounded-wall network can be detected for $U(1)_R$ breaking scale from $10^{12}$ to $10^{15}$ GeV 
in forthcoming interferometers including LISA and Einstein Telescope, 
with a distinctive $f^{3}$ slope and inflexion in the frequency range between microhertz and hertz.    
}

\begin{document}
\maketitle
\flushbottom

\section{Introduction}

After the discovery of the Higgs boson \cite{ATLAS:2012yve,CMS:2012qbp}, one of the next goals for the collider experiments is to investigate the properties of the Higgs boson. The 2012 data has already shown an excess at 95 GeV \cite{CMS:2018cyk} and next generation collider experiments have been proposed to measure the Higgs boson interactions precisely \cite{ILC:2013jhg,An:2018dwb,Cepeda:2019klc}. 
While revealing the nature of the standard model (SM) Higgs boson, indication of new physics beyond the standard model (BSM) may also appear.

One of the most attractive models that can be tested by measuring the properties of the Higgs boson is the two-Higgs-doublet model (2HDM) \cite{Lee:1973iz}.
In 2HDM, a serious potential problem is the tree-level flavour-changing neutral currents (FCNCs). 
The necessary and sufficient condition to avoid FCNC is that all fermions with the same quantum numbers couple to the same Higgs doublet \cite{Glashow:1976nt,Paschos:1976ay}, leading to limited types of the 2HDM \cite{Branco:2011iw}.
A simple way to realise a specific type of 2HDM is to impose an extra symmetry of the model. 
The most popular possibility is to impose a $Z_2$ discrete symmetry, which can be used to construct many different types of 2HDM \cite{Ivanov:2017dad}.
However, as the SM is a gauged theory, it is more interesting and technically elegant to consider a gauge extension of the model symmetry. 

In this work, we consider a gauged $U(1)$ extension of the SM gauge group as the origin of natural flavour conservation \cite{Ko:2012hd,Ko:2013zsa,Ko:2014uka,Campos:2017dgc}. 
Furthermore, in order to explain the neutrino oscillation, we include right-handed neutrinos (RHNs), which are responsible for generating the light neutrino masses through the type-I seesaw mechanism \cite{Minkowski:1977sc,Yanagida:1979as,GellMann1979,Glashow:1979nm,Mohapatra:1979ia}. Assuming the SM fermion content is only extended by three right-handed neutrinos,\footnote{Other types of 2HDM can be implemented by adding vector-like fermions, see Ref.~\cite{Ko:2012hd}.} in the case of the type-I 2HDM (but not other types), it is possible to control 
FCNCs by an anomaly free gauged $U(1)$ symmetry. In general, type-I 2HDMs have a rich lab-based phenomenology \cite{Arhrib:2016wpw,Arhrib:2017uon,Enberg:2018nfv,Wang:2021pxc,Wang:2021zjp,Moretti:2022fot,Moretti:2022okg,Mondal:2023wib,Arhrib:2023apw,Liu:2024azc,Chen:2019pkq,Han:2021udl,Han:2020lta,Bian:2017gxg,Chen:2020soj,DelleRose:2017xil}. Here we shall focus on the implications of the gauged $U(1)$ type-I 2HDM for gravitational wave signatures.

In addition to the Higgs doublets, a complex scalar singlet is also added to the scalar sector, which gives the right-handed neutrinos Majorana masses at the seesaw scale. Motivated by experimental constraints \cite{ATLAS:2022vkf,CMS:2022dwd}, we consider the so-called ``alignment limit'', where one of the scalar mass eigenstates recovers the properties of the SM Higgs boson.
The vacuum expectation value (VEV) of the complex scalar singlet $\phi$ breaks the $U(1)_R$ symmetry, under which only the right-handed fermions are charged, to a residual $Z_2$ symmetry. 
Later, the $Z_2$ symmetry is further broken by the VEVs of Higgs doublets $\Phi_i$ at the electroweak (EW) scale,
\begin{align}
    U(1)_R \stackrel{\langle \phi \rangle}{\xrightarrow{\hspace{0.8cm}}} Z_2 \stackrel{\langle \Phi_i \rangle}{\xrightarrow{\hspace{0.8cm}}} I. \label{UtoZ2}
\end{align}

A natural outcome of this model is that the $U(1)_R$ symmetry is spontaneously broken to a residual $Z_2$ symmetry at a high scale, motivated by requiring heavy Majorana mass of the RHNs. 
This leads to the creation of a cosmic string network \cite{Kibble:1976sj}.
This network forms closed loops, and the vibrating loops loose their energy via emitting gravitational waves \cite{Vachaspati:1984gt}. Later, when the Higgs doublets get a VEV around the EW scale, the $Z_2$ symmetry is spontaneously broken, leading to the creation of domain walls \cite{Kibble:1976sj}.
These domain walls are bounded by the pre-existing cosmic string network, creating a hybrid network of ``domain walls bounded by cosmic strings'' \cite{Kibble:1982dd, Everett:1982nm, Lazarides:2023iim}.\footnote{Such topological defects, also dubbed as ``Kibble-Lazarides-Shafi (KLS) walls'', have been found in the polar-distorted B phase of superfluid ${}^3$He \cite{Makinen:2018ltj}.}

The appearance of topological defects in a symmetry breaking chain can be understood from the homotopy groups of the quotient space \cite{Kibble:1976sj, Preskill:1992ck, Vilenkin:2000jqa}. For the breaking chain $G \rightarrow H \rightarrow I$, the first stage breaking $G\rightarrow H$ would produce a cosmic string defect if the first homotopy group of the quotient space $G/H$ is nontrivial. In the present case, the first stage breaking is $U(1)_R \rightarrow Z_2$, where the first homotopy group of the quotient space is nontrivial, $\pi_1 (U(1)_R/Z_2) \simeq \pi_1 (U(1)_R) \simeq Z$, hence this breaking creates a cosmic string network. For the second breaking $Z_2 \rightarrow I$, the zeroth homotopy group of the quotient space is nontrivial, $\pi_0(Z_2) = Z_2$, hence this breaking leads to the appearance of domain walls. One might be tempted to apply the theorem $\pi_1(G/H) = \pi_0(H)$ for a general breaking chain $G \rightarrow H$ here, but this requires $G$ to be a connected and simply-connected Lie group; in the present case $G \equiv U(1)$ is connected but not simply-connected. To summarize, for the breaking chain $U(1)_R \to Z_2 \to I$, successive stages of symmetry breaking produce cosmic strings and then domain walls. Since the cosmic string network already exists when the domain walls appear, walls get attached to the strings. In our case, the strings have already formed loops, so the domain walls appear bounded by strings.

{The existence of domain walls does not change the stability of the cosmic string but affect the dynamics of the string loops.}
The interaction between the surface tension of the wall on the oscillating string loop accelerates its decay, resulting in an earlier collapse of the string network that loses energy by emitting gravitational waves. 
The net observational effect is the modification of the gravitational wave spectrum at lower frequencies that correspond to the late stage evolution of the defect network, while the high frequency spectrum remain flat, as is also seen in the case of pure strings in the absence of any intermediate symmetry breaking. 
In particular, the hybrid network features a signal with $f^{3}$ slope in the infrared frequencies, quite distinguishable from the pure string case \cite{Dunsky:2021tih, Maji:2023fba, Lazarides:2023ksx}. 
The amplitude of the high frequency flat spectrum conveys information about the $U(1)_R$ breaking scale, while the transition frequency where the infrared tail starts depends on the intermediate $Z_2$ breaking scale. 
Since the latter is quite constrained around the EW scale in the present model, we have a sharp prediction for GW signals with distinct spectral slopes in the microHz -- Hz band.  

{The paper is structured as follows.} In Sec.~\ref{sec:2HDM}, we discuss the anomaly-free $U(1)$ extension of the 2HDM with right-handed neutrinos. Section~\ref{sec:hybrid} investigates the emergence of the hybrid topological defect, and the associated GW signal in this model. We report our main results in Sec.~\ref{sec:GW} and finally conclude in Sec.~\ref{sec:conclusion}.

\section{Type-I 2HDM extended with a gauged $U(1)$ symmetry \label{sec:2HDM}}

We consider a gauged $U(1)$ extension of the SM gauge group in 2HDM with three right-handed neutrinos.
As a fundamental principle of extending the gauge group, the chiral anomaly has to be cancelled. 
Suppose the charge of a particle $i$ under the extra $U(1)$ symmetry is $X_i$. 
To achieve anomaly cancellation, the charges of other fermions have to satisfy \cite{Ko:2012hd,Ko:2013zsa,Ko:2014uka,Campos:2017dgc}:
\begin{eqnarray}
X_q = \frac12 (X_d + X_u)\,,\,  X_l = -\frac32 (X_d + X_u)\,,\,  X_e = -X_d - 2 X_u\,,\,  X_n = -2 X_d - X_u\,.
\end{eqnarray}
$X_q$, $X_u$, $X_d$, $X_l$, $X_e$ and $X_n$ are the $U(1)$ charges of left-handed quark doublet, right-handed up-type quark, right-handed down-type quark, left-handed lepton doublet, right-handed charged lepton and right-handed neutrino, respectively. 
Such a solution automatically leads to $X_u-X_q=X_q-X_d=X_l-X_e=X_n-X_l$, meaning that the $U(1)$ charge of the Higgs doublet coupling to the fermions is enforced to be $(X_u-X_d)/2$. 
In order to avoid FCNC, the Higgs doublets must have different quantum number, which ensures that only one of the Higgs doublets, $\Phi_2$, can couple to the fermions while the other, $\Phi_1$, is fermiophobic.
The interaction between the fermiophobic Higgs doublet and the fermions is forbidden as long as its charge is different from $(X_u-X_d)/2$.
The type of 2HDM with such a structure in the Yukawa sector is the well-known type-I 2HDM.

The $U(1)$ charge of the right-handed neutrinos forbids their Majorana masses. 
In order to realise the seesaw mechanism, a scalar singlet $\phi$ is included in the model, which couples to the right-handed neutrinos through $\phi \overline{N^c} N$. 
The VEV of $\phi$ breaks the extra $U(1)$ symmetry and gives mass to the right-handed neutrinos.
As the charge of $\Phi_1$ is not constrained due to the absence of its interaction with fermions, it is possible for the scalar doublets to couple to the scalar singlet through interaction of the type $(\Phi_1^\dagger \Phi_2 \phi + \mathrm{h.c})$. 
Such kind of possibility has been discussed in Ref.~\cite{Campos:2017dgc} with many concrete examples.

In general, when the extra $U(1)$ symmetry is broken by the VEV of $\phi$, there might be a residual discrete symmetry. 
The exact type of the residual symmetry is determined by the charge of $\phi$ and the minimal charge unit of the fields.
For example, in the case that $X_u=1$ and $X_d=0$, the $U(1)$ charge of $\phi$ is 2 while the minimal charge unit of the fields is $1/2$. 
After $\phi$ gets a VEV, the $U(1)$ symmetry would be broken into a $Z_4$. 
Here, as the gravitational wave phenomena associated with the $Z_2$ domain wall are the most well-understood, we choose $X_u=1$ and $X_d=-1$ for simplicity. 
In such a case, the right-handed fermions are charged under the $U(1)$ while the left-handed fermions are not, hence we will call this $U(1)_R$.\footnote{Not to be confused with the continuous $R$-symmetry in the context of supersymmetry, or the right-handed $U(1)$ symmetry arising from left-right symmetric models.}
The particle content is listed in Table~\ref{tab:symmetry}.
After $\phi$ gains a VEV, the $U(1)_R$ is broken into a $Z_2$ symmetry, which is the simplest cyclic group.
\begin{table}[t!]
\renewcommand\arraystretch{1.0}
\centering
\begin{tabularx}{0.75\textwidth}{c c c c c c c c c c c c c }
\toprule 
& ${u_R}_\beta$ & ${d_R}_\beta$ & ${Q}_\alpha$ 
& ${L}_\alpha$ & ${e_R}_\beta$ & $N_{R\beta}$ 
& $\Phi_{2}$ & $\Phi_{1}$ & $\phi$ \\ \midrule 
$SU(2)_L$ & {\bf 1} & {\bf 1} & {\bf 2} & {\bf 2} & {\bf 1} & {\bf 1} & {\bf 2} & {\bf 2} & {\bf 1} \\ 
& & & & & & & & & \\ [-1em]
$U(1)_Y$ & $\frac{2}{3}$ & $-\frac{1}{3}$ & $\frac{1}{6}$ & $-\frac{1}{2}$ & $-1$ & 0 & $\frac{1}{2}$ & $\frac{1}{2}$ & 0 \\[2pt] 
& & & & & & & & & \\ [-1em]
$U(1)_R$ & $1$ & $-1$ & $0$ & $0$ &  $-1$ & $1$ & $1$ & $-1$ & $-2$ \\ [2pt] 
residual $Z_2$ & $-$ & $-$ & $+$ & $+$ & $-$ & $-$ & $-$ & $-$ & $+$\\[2pt] \bottomrule 
\end{tabularx}
\caption{\label{tab:symmetry} Field content of the type-I 2HDM with gauged $U(1)_R$ symmetry and its residual $Z_2$ subgroup.}
\end{table}

With the particle content in Table~\ref{tab:symmetry}, the allowed fermion Yukawa interactions are 
\begin{eqnarray}
\mathcal{L}_Y \supset Y_u \overline{Q}\tilde{\Phi}_2 u_R + Y_d \overline{Q} \Phi_2 d_R + Y_e \overline{L}\Phi_2 e_R + Y_N \overline{L} \tilde{\Phi}_2 N_R + y_N\phi N_R N_R+ \text{h.c.},
\end{eqnarray} 
where $\tilde{\Phi}_2=-i\sigma_2\Phi_2^*$.
Between the Higgs doublets, only $\Phi_2$ is allowed to couple to the fermions.
The most general scalar potential allowed by the $SU(2)_L\times U(1)_Y \times U(1)_R$ symmetry is 
\begin{eqnarray}
V(\Phi_1,\,\Phi_2,\,\phi) &=& 
m_{11}^2\Phi_1^\dagger \Phi_1 
+ m_{22}^2\Phi_2^\dagger \Phi_2 
+ m_{\phi}^2|\phi|^2 
- m_{12} \left(\Phi_1^\dagger \Phi_2 \phi + \mathrm{h.c} \right)
\nonumber \\&& 
+ \frac{\lambda_{1}}{2} (\Phi_1^\dagger \Phi_1)^2 
+ \frac{\lambda_{2}}{2} (\Phi_2^\dagger \Phi_2)^2 + \lambda_{3} (\Phi_1^\dagger \Phi_1)(\Phi_2^\dagger \Phi_2) + \lambda_{4} (\Phi_1^\dagger \Phi_2)(\Phi_2^\dagger \Phi_1) 
\nonumber \\&& 
+ \lambda_{1\phi} (\Phi_1^\dagger \Phi_1) |\phi|^2 
+ \lambda_{2\phi} (\Phi_2^\dagger \Phi_2) |\phi|^2 
+ \lambda_{\phi} |\phi|^4 \,.
\end{eqnarray} 
The potential is CP conserving as the phase of $m_{12}$ can be eliminated through rephasing the scalar fields and all the other parameters are required to be real by hermiticity. 
The $m_{12}$ term breaks the normal $B-L$ symmetry explicitly and thus there is no spontaneous $U(1)_{B-L}$ symmetry breaking.\footnote{Different from Ref.~\cite{Campos:2017dgc}, the $B-L$ symmetry refers to the one under which both of the Higgs doublets are uncharged.}
Despite the absence of a strict $B-L$ symmetry, the Majorana mass of the right-handed neutrinos is still protected by the $U(1)_R$ symmetry.
After $\phi$ gets a VEV $\langle\phi\rangle = v_{\rm{M}}$, the potential of the Higgs doublets becomes 
\begin{eqnarray}
V(\Phi_1,\,\Phi_2) &=& 
\tilde{m}_{11}^2\Phi_1^\dagger \Phi_1 
+ \tilde{m}_{22}^2\Phi_2^\dagger \Phi_2 
- \tilde{m}_{12}^2 \left(\Phi_1^\dagger \Phi_2 + \mathrm{h.c} \right)
\nonumber \\&& 
+ \frac{\lambda_{1}}{2} (\Phi_1^\dagger \Phi_1)^2 
+ \frac{\lambda_{2}}{2} (\Phi_2^\dagger \Phi_2)^2 
+ \lambda_{3} (\Phi_1^\dagger \Phi_1)(\Phi_2^\dagger \Phi_2) 
+ \lambda_{4} (\Phi_1^\dagger \Phi_2)(\Phi_2^\dagger \Phi_1)\,,
\label{eq:potential} 
\end{eqnarray} 
where $\tilde{m}_{11}^2 = m_{11}^2 + \lambda_{1\phi} v_{\rm{M}}^2$, $\tilde{m}_{22}^2 = m_{22}^2 + \lambda_{2\phi} v_{\rm{M}}^2$ and $\tilde{m}_{12}^2 = m_{12} v_{\rm{M}}$. 
Unlike the most general CP-conserving potential for 2HDMs, the interaction $\frac{\lambda_5}{2} ((\Phi_1^\dagger \Phi_2)^2 + \mathrm{h.c})$ is still absent due to the $U(1)_R$ symmetry.

The most general 2HDM admits CP-breaking vacua, where the VEVs of the Higgs doublets have a relative phase.
However, as all the parameters in the scalar potential in Eq.~\eqref{eq:potential} are real, the vacua are always real and can be expressed as $\langle\Phi_i\rangle = v_i/\sqrt2$, where $\sqrt{v_1^2 + v_2^2} = v_{\rm{SM}} \sim 246$ GeV.\footnote{A relative phase $\varphi$ only affects the potential by a factor $\cos\varphi$ in the $\tilde{m}_{12}^2$ term. Since $\tilde{m}_{12}^2$ is real, minimising the potential leads to $\cos\varphi=1$ for $\tilde{m}_{12}^2>0$ and $\cos\varphi=-1$ for $\tilde{m}_{12}^2<0$.  } 
The ratio $v_2/v_1$ is defined as $\tan\beta$.
Rotating the Higgs fields by an angle $\beta$, we obtain the so-called ``Higgs basis'', where only one of the doublets gets a nonzero VEV.
Another popular basis for 2HDM is the ``mass eigenbasis'', found by a rotating the fields by an angle $\alpha$.
However, the properties of SM Higgs are not recovered by any field in either of these choices.
Instead, a combination defined as 
\begin{align}
    H_{\rm SM} = H\cos{(\beta-\alpha)} + h\sin{(\beta-\alpha)}
    \label{Hsm}
\end{align}
recovers both the gauge interaction and Yukawa interaction of the SM Higgs boson, where $H = (\Phi_1 \cos\alpha + \Phi_2\sin\alpha)$ and $h = (\Phi_2\cos\alpha - \Phi_1 \sin\alpha)$ are the two mass eigenstates. For details see Appendix \ref{app:scalars}.

\subsection{The alignment limit and the domain wall solution}

As the properties of the Higgs boson discovered in 2012 are consistent with the predication of the SM at around $2\sigma$ level \cite{ATLAS:2022vkf,CMS:2022dwd}, we first consider the ``alignment limit''  \cite{Gunion:2002zf,Haber:2022swy,BhupalDev:2014bir}, where one of the Higgs mass eigenstates recovers the properties of the SM Higgs boson.
This is achieved, for example, by imposing the relation $\alpha = \beta - \pi/2$, which makes $h$ the SM-like Higgs boson, i.e. $h=H_{\rm SM}$ in Eq.~\eqref{Hsm}.

\begin{figure}[t!]
\begin{center}
\subfloat[]{\includegraphics[height=0.36\textwidth]{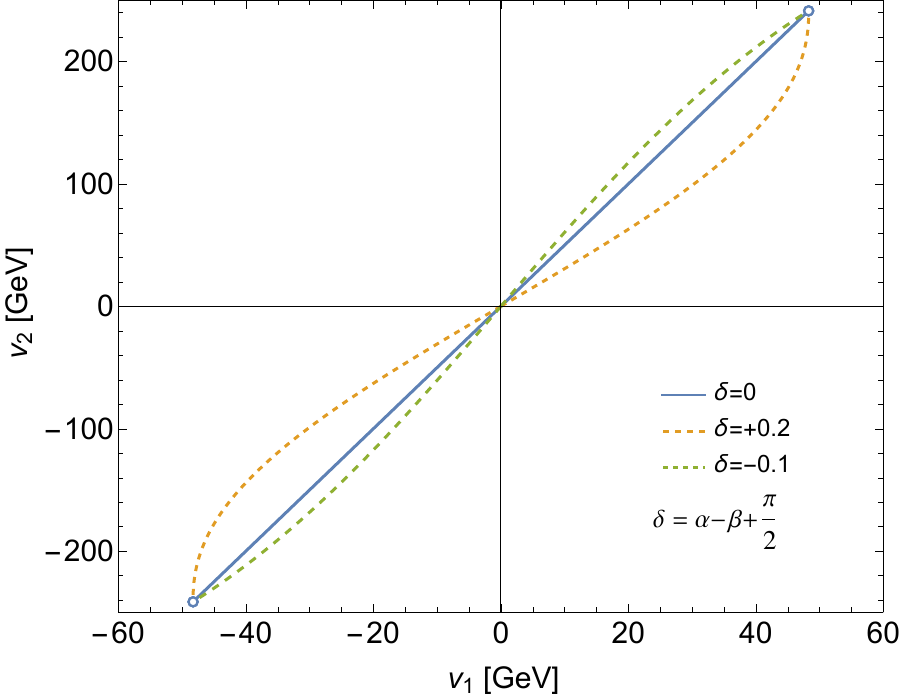}\label{fig:path}}
\quad
\subfloat[]{\includegraphics[height=0.376\textwidth]{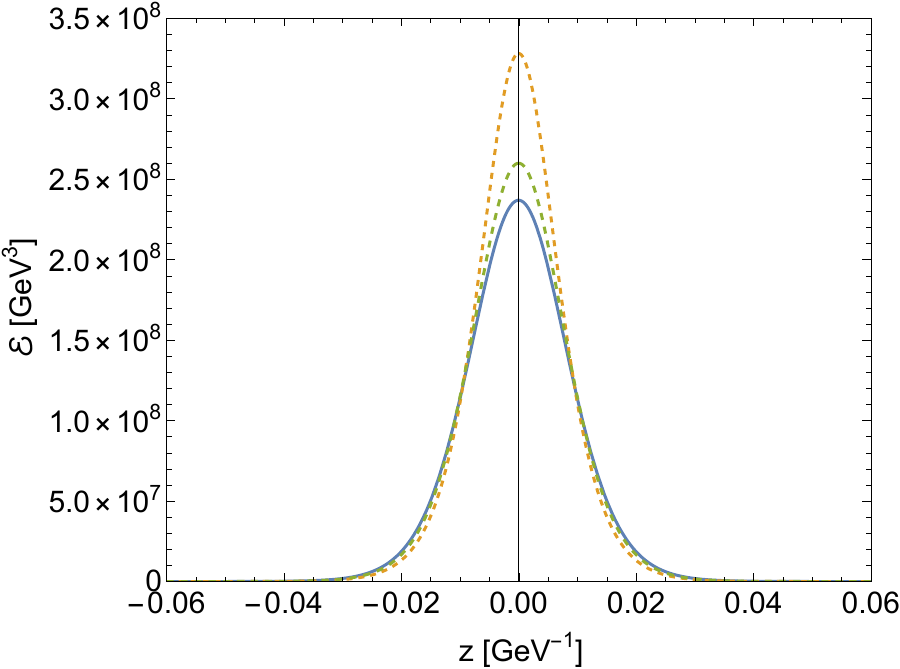}\label{fig:En}}
\caption{Path in the field space (left) and energy density across the domain wall (right). The energy density $\mathcal{E}$ for $\delta = + 0.2$ and $\delta = - 0.1$ are $5.75 \times 10^6 \, \text{GeV}^3$ and $5.25 \times 10^6 \, \text{GeV}^3$, respectively. }
\end{center}
\end{figure}

The scalar potential is symmetric under the surviving $Z_2$ in Eq.~\eqref{UtoZ2} for $\Phi_i \to -\Phi_i$. When the Higgs doublets get VEV $v_i$, this symmetry is spontaneously broken, resulting in regions of space dubbed as ``domains'' randomly adopting the $+v_i$ or $-v_i$ vacua. These regions are separated by walls with a static planar configuration, taken to be perpendicular to the $z$ direction without loss of generality. The field equation for the domain walls can be expressed as \cite{Saikawa:2017hiv}
\begin{eqnarray}
\frac{d^2\Phi_i}{dz^2} = \frac{\partial V}{\partial \Phi_i}\,,\quad i=1,2\,,
\end{eqnarray}
with the boundary conditions $\Phi_i (z \to \pm \infty) = \pm v_i$ at two different vacua. 
For a general potential with multiple scalar fields, one needs to consider the parallel and perpendicular parts of the equations of motion, which are 
\begin{eqnarray}
\frac{d^2x}{dz^2} &=& \frac{d V[\vec{\Phi} (x)]}{d x}\,, \label{eq:parallel}\\
\frac{d^2\vec{\Phi}}{dx^2}\left(\frac{dx}{dz}\right)^2 &=& \nabla_{\perp} V(\vec{\Phi})\,, \label{eq:perpendicular}
\end{eqnarray}
and apply the path deformation method \cite{Wainwright:2011kj,Chen:2020soj} to find the solution. 
However, in the alignment limit, the correct path is simply the straight line connecting the two minima in the field space (the blue path in Fig.~\ref{fig:path}). 
Indeed, once we choose the path to be $\Phi_1=x\cos\beta$ and $\Phi_2=x\sin\beta$, Eq.~\eqref{eq:perpendicular} is automatically satisfied regardless of the relation between $x$ and $z$ (see Appendix \ref{app:basis} for details). 
Then the problem is reduced to finding the domain wall solution for Eq.~\eqref{eq:parallel}, which is simply
\begin{eqnarray} \label{eq:dw}
\frac{d^2x}{dz^2} &=& \frac{m_h^2}{2 v_{\rm SM}^2} x (x^2 -  v_{\rm SM}^2) = \lambda x (x^2 -  v_{\rm SM}^2)\,
\end{eqnarray}
with $m_h=125$ GeV the mass of the SM-like Higgs $h$ and $\lambda\equiv m_h^2/2 v_{\rm SM}^2$ the SM Higgs quartic coupling. 
Eq.~\eqref{eq:dw} admits a well-known analytical solution \cite{Kolb:1988aj} 
\begin{eqnarray}
x(z) = v_{\rm SM} \tanh{\left(z/\Delta\right)}\,,
\end{eqnarray}
where $\Delta = \sqrt{2/\lambda}/v_{\rm SM}$ is the thickness of the wall. 
Then it is straightforward to obtain the profiles of the Higgs doublet fields
\begin{eqnarray}
\Phi_1(z) = v_{\rm SM} \tanh{\left(z/\Delta\right)} \cos\beta 
\quad \text{and} \quad
\Phi_2(z) = v_{\rm SM} \tanh{\left(z/\Delta\right)} \sin\beta \,.
\end{eqnarray}
The surface tension, or the surface energy density, of the wall can be obtained by integrating the energy density over the space coordinate $z$, which gives 
\begin{eqnarray}
\mathcal{E} \approx \int_{-\infty}^{\infty} dz \sum_i \frac{1}{2}\left(\frac{\partial \Phi_i}{\partial z}\right)^2 =  \frac{2}{3}\sqrt{2\lambda} v_{\rm SM}^3\,,
\end{eqnarray}
which is around $5.04 \times 10^6 \, \text{GeV}^3$ given the SM Higgs VEV and self coupling.

\subsection{Deviation from the alignment limit}

\begin{figure}
    \centering
    \includegraphics[width=0.5\textwidth]{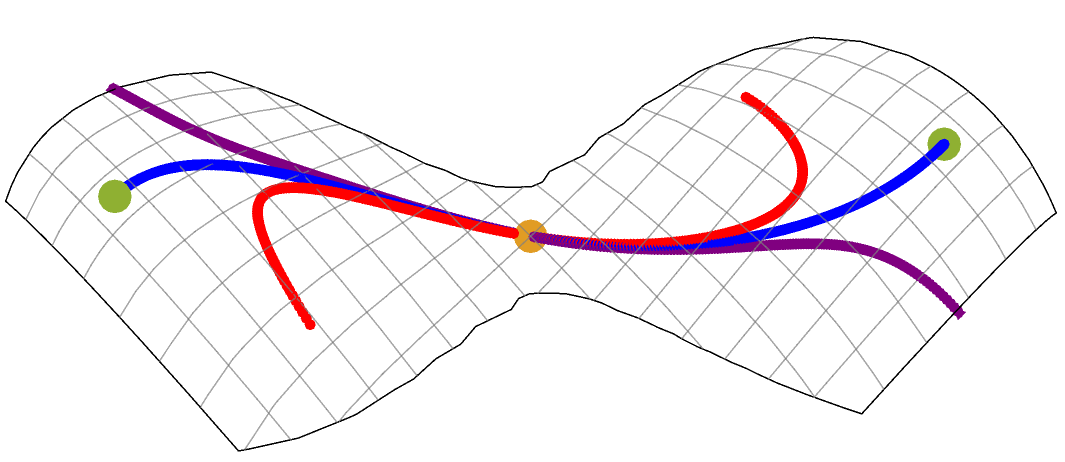}
    \caption{Paths found by the overshoot/undershoot method. The blue path exactly connects the two minima (green points).
    The red and purple lines show two paths with the wrong direction.
    The yellow point is the symmetry center.}
    \label{fig:shoot}
\end{figure}

Although the experiments indicate the alignment limit is a good approximation, a deviation has not been ruled out yet. 
Allowing a deviation from the alignment limit, the domain wall solution would no longer follow the straight line between the two equivalent minima (see Appendix \ref{app:basis} for details). 
The domain wall solution can be found by the path deformation method (see for example \cite{Wainwright:2011kj,Chen:2020soj}), or alternatively, by the overshoot/undershoot method (see for example \cite{Apreda:2001us, Wainwright:2011kj}).
As the scalar potential possesses a central symmetry, the domain wall path has to pass the center, which is the origin. 
The magnitude of the field profile derivative can be determined everywhere in the field space as $|d\vec{\phi}/dz|=\sqrt{2\Delta V}$ in the spirit of the overshoot/undershoot method. 
Eventually the problem is reduced to finding the direction of the path at the origin.
The path exactly connect the two minima for the correct direction, while the wrong direction never pass the minima without any ends, as shown in Fig.~\ref{fig:shoot}.
Using both methods, we solve the coupled system of field equations \eqref{eq:parallel} and \eqref{eq:perpendicular}, and explore how much a deviation from the SM-like Higgs limit can change the domain wall energy density. 

An illustrative example of this is shown in Fig.~\ref{fig:path}. 
After the spontaneous symmetry breaking, there are five physical scalar fields, namely, two neutral scalars (one of which, $h$, aligns with the SM Higgs in the alignment limit), a charged scalar, and a pseudoscalar.
For the present example, we choose a benchmark point {where the masses of the neutral scalar different from the SM-like Higgs, the charged scalar and the pseudoscalar are $410$ GeV, $380$ GeV and $400$ GeV, respectively, while $\tan\beta=5$.
Details on the relations between the masses of physical scalar fields and model parameters are discussed in Appendix \ref{app:scalars}.} The path in the field space for the domain wall solution follows a straight line in the perfect alignment limit, as shown in Fig.~\ref{fig:path}.
On the other hand, allowing a deviation $\delta$ from the alignment limit, i.e., $\alpha = \beta -\pi/2 +\delta$, the path changes.
In particular, we consider two extremes of the deviation angle $\delta = + 0.2$ and $\delta = - 0.1$ which would reduce the coupling between $h$ and gauge bosons by around 2\% and 0.5\%, respectively.\footnote{A deviation $\delta$ from the alignment limit changes both the gauge interaction ($h W^\mu W_\mu$) and the Yukawa interaction ($h \overline{t} t$) of $h$ by a factor $\cos\delta$. $\delta = +0.2$ and $\delta = -0.1$ correspond to $\cos\delta\simeq 0.98$ and $\cos\delta\simeq 0.995$, respectively.} 
Such deviations represent the limit allowed by the experimental constraints \cite{CMS:2022dwd,ATLAS:2022vkf} while preserving the stability of the vacua.\footnote{A deviation $\delta = -0.2$ leads to metastable vacua for the benchmark case.} 
The energy density in the direction perpendicular to the wall is presented in Fig.~\ref{fig:En}.
In the presence of the deviation, the energy at the center of the domain wall becomes larger, leading to an increase in the total surface energy density. By exploring the domain wall solutions in the model, we have verified that the allowed deviation from the alignment limit has limited influence on the domain wall tension, and thus has unnoticeable impact on the gravitational wave signals. In the rest of the paper, we will adopt the perfect alignment limit. 

\section{Domain walls bounded by cosmic strings} \label{sec:hybrid}

In the type-I 2HDM extended by a gauged $U(1)_R$, the symmetry breaking pattern is as follows. 
The scalar $\phi$ gets a nonzero VEV at a high scale and gives large mass to the right-handed neutrino. 
This spontaneously breaks the $U(1)_R$ symmetry and creates a cosmic string network \cite{Kibble:1976sj}. 
We assume that this happens in a radiation dominated Universe after inflation has ended. 
This ensures that the string network is not diluted by inflation.\footnote{Unlike other topological defects, string networks are not completely diluted away by inflation, and can even regrow to generate observable gravitational wave signatures. See \cite{Guedes:2018afo,Cui:2019kkd,Ferrer:2023uwz} for details. 
In case that the string is completely diluted, hybrid topological defects can still appear through nucleation of holes bounded by strings on the domain wall \cite{Kibble:1982ae,Kibble:1982dd}. However, such topological defects can only lead to observable gravitational wave for quasi-degenerate string scale and wall scale, with inflation scale in the middle. }

Besides, the strings naturally interact with the right-handed neutrinos due to their coupling with $\phi$, which may lead to interesting phenomena.
In general, when the scalar field that breaks the $U(1)$ symmetry couples to other fields, the resulting strings may carry neutral or charged current \cite{Witten:1984eb}. 
For example, the strings can carry a neutral current due to the RHNs coupling to the scalar. 
Moreover, if the model is embedded into higher groups, the cosmic strings can also carry electric currents (known as superconducting strings) \cite{Afzal:2023kqs}. 
Such strings carrying currents lead to a GW spectrum different from the standard one over a wide frequency range due to the presence of the fermion zero mode, as analysed in detail in Refs.~\cite{Sousa:2020sxs,Auclair:2022ylu}. 
In our model, due to the presence of the charge of the RHN (under which it interacts with the scalar $\phi$), the correlation length of the strings is modified since the loops formed in the radiation era show non-trivial evolution. 
Such a study will involve treating walls bounded by strings which carry current and need detailed simulation and exploration of the scaling regime, which is beyond the scope of the present paper and will be taken up later in future publications.

Because of the $U(1)_R$ charge assignment of the fields, as given in Table \ref{tab:symmetry}, the model has a residual $Z_2$ symmetry at the $\phi$ vacuum, under which both Higgs doublets $\Phi_1$ and $\Phi_2$ are odd. When either of them gets a nonzero VEV, the $Z_2$ symmetry is spontaneously broken, leading to the creation of a domain wall network \cite{Kibble:1976sj,Battye:2011jj,Battye:2020jeu,Battye:2020sxy}. 

\begin{figure}[t!]
    \centering
    \subfloat[][Before $Z_2$ symmetry breaking. \label{fig2a}]{\includegraphics[height=0.22\textwidth]{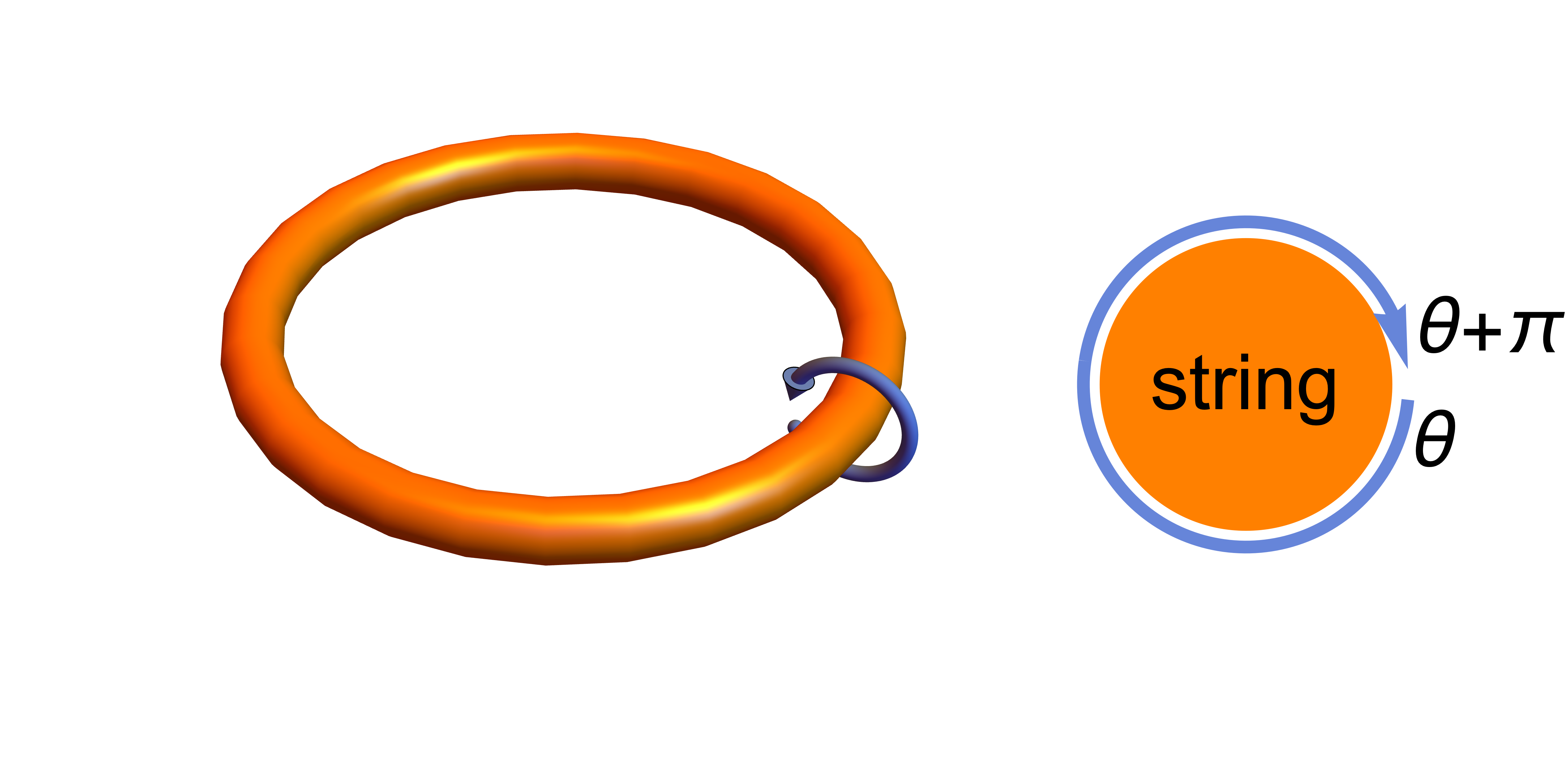}}
     \subfloat[][After $Z_2$ symmetry breaking. \label{fig2b}]{\includegraphics[height=0.22\textwidth]{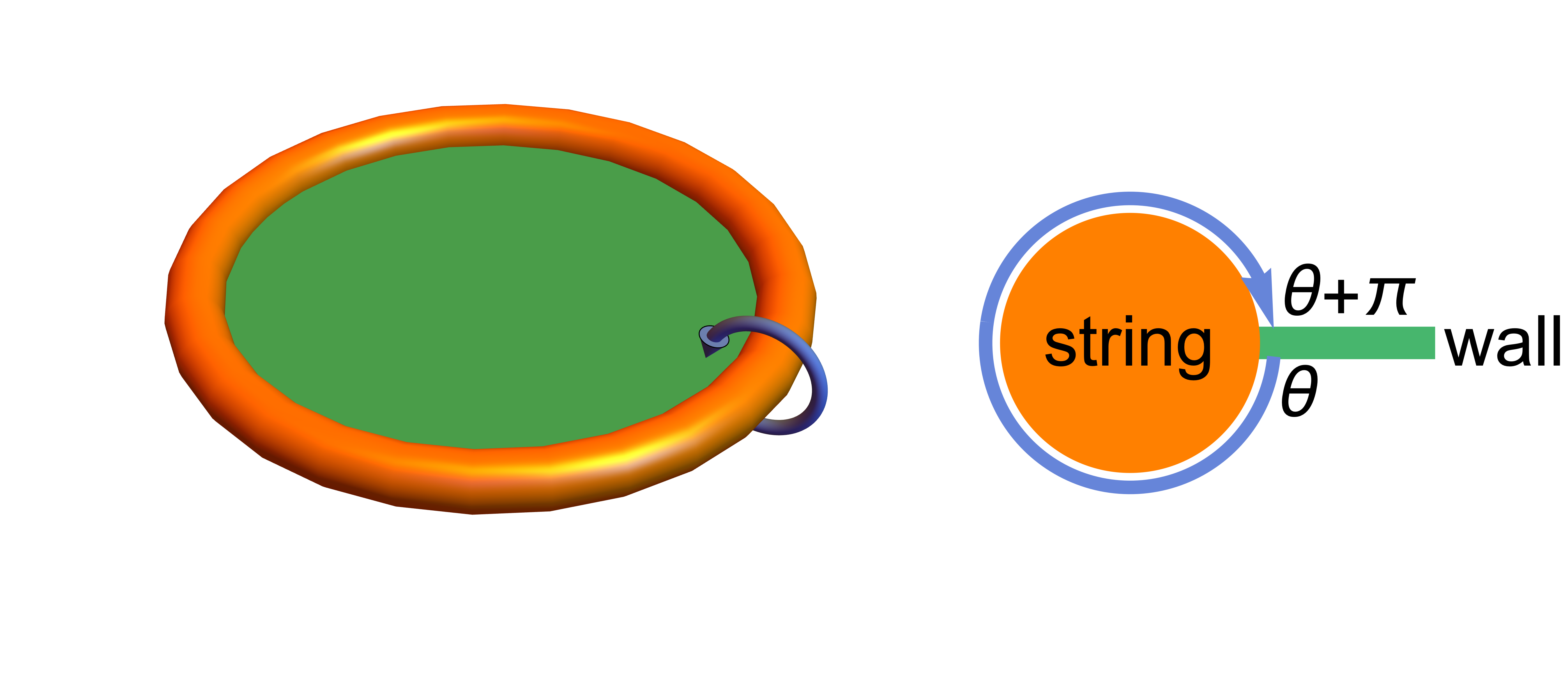}}
    \caption{Illustration of the appearance of the hybrid topological defect. The spontaneous breaking of $U(1)_R$ into a residual $Z_2$ symmetry creates a string network. When the $Z_2$ symmetry is spontaneously broken later, domain walls appear which fill up the space between the string loops.}
    \label{fig:wall-string}
\end{figure}

While the creation of cosmic strings and domain walls from their respective symmetry breaking have been widely studied in the context of various BSM models, the situation above is somewhat different. 
Here the $Z_2$ is not a symmetry of the original Lagrangian, but is a residual symmetry that occurs at an intermediate stage in the symmetry breaking chain. 
When the $U(1)_R$ symmetry is broken, the VEV of the scalar field with $U(1)_R$ charge 2 can have a phase $\theta(x)$ that can vary between $0$ and $\pi$ with the spacetime. 
The same applies to the other fields in the theory. 
After the $U(1)_R$ symmetry breaking, the phase changes from $\theta(x)$ to $\theta(x)+\pi$ around a string, as shown in Fig.~\ref{fig2a}.
However, those fields with $U(1)_R$ charge 1 are not invariant when $\theta(x)$ changes to $\theta(x)+\pi$.
Instead, there is a sign change in those fields, leading to a residual $Z_2$ symmetry.
As such, the created domain walls after $Z_2$ breaking are not infinite planar walls, rather circular walls that fill in the space between string loops, as shown in Fig.~\ref{fig2b}.\footnote{We focus on string loops. In case that the wall is formed between long strings, the wall would promote the loop formation \cite{Leblond:2009fq}.} This creates a hybrid topological defect dubbed as ``walls bounded by strings'' \cite{Kibble:1982dd, Everett:1982nm}.

A detailed description of the evolution of this hybrid topological defect can be found in Ref.~\cite{Dunsky:2021tih}. 
Here we will briefly discuss the relevant physics and the observational signatures of the defect.

A useful parameter that dictates the evolution of the defect is defined as $R_c \equiv \mu / \mathcal{E}$, where $\mu \sim v_{\rm M}^2$ is the string tension, $v_{\rm M}$ being the $U(1)_R$ breaking scale, and $\mathcal{E} \sim v_{\rm{SM}}^3$ is the surface energy density of the domain wall, $v_{\rm{SM}}$ being the $Z_2$ breaking scale, and assuming a $\mathcal{O}(1)$ coupling constant of the scalar field breaking $Z_2$. After the $U(1)_R$ symmetry breaking, horizon size long strings are created that intersect and form string loops. These circular loops oscillate non-relativistically  and lose their energy by emitting gravitational waves. When domain walls are formed at $t_{\rm DW} \approx M_{\rm Pl} C / v_{\rm{SM}}^2$, where $C = (8\pi^3 g_\star/ 90)^{-1/2}$, and fill up the space between the string loops, they tend to make the string motion ultra-relativistic as long as the loop size is above $R_c$. This makes the wall-string network collapse earlier than the case of pure string loops. If $t_{\rm DW} < R_c$, walls do not dominate the string dynamics immediately after creation. As such, the hybrid network becomes ultra-relativistic and collapses after some time, $t_\star \sim R_c$. On the other hand, if $t_{\rm DW} > R_c$, the hybrid network becomes ultra-relativistic as soon as the walls are created and the network collapses at $t_\star \sim t_{\rm DW}$. We can, therefore, define the time for the collapse of the hybrid network as $t_\star \equiv \max(R_c, t_{\rm DW})$ \cite{Martin:1996ea}.

For a loop of radius $R$ with circular length $l = 2\pi R$, the rate of energy loss is given by
\begin{align}
    \frac{dE}{dt} = -\Gamma(l) G\mu^2,
\end{align}
where $G$ is the Newton's constant. The function $\Gamma(l) \approx \Gamma_s = 50$ when $l \ll 2\pi R_c$ in the pure string limit, and $\Gamma(l) = 3.7(l/(2\pi R_c))^2$ for $l \gg 2\pi R_c$ in the pure wall limit. The behavior around $l \sim 2\pi R_c$ can be approximated using a smooth interpolation function, as shown in Ref.~\cite{Dunsky:2021tih}.

Because of the energy loss, loops forming at time $t_k$ with an initial size $l_k = \alpha t_k$ slowly decrease in size with time. Here $\alpha \approx 0.1$ is the ratio between loop formation length and horizon size, and its value is determined from simulations \cite{Blanco-Pillado:2017oxo, Blanco-Pillado:2013qja}. For the domain walls bounded by strings, the size of a string loop emitting gravitational waves at time $\tilde{t}$ is $\tilde{l} = 2\pi \tilde{R}$, and we have
\begin{align}
    G\mu (\tilde{t}-t_k) = \int_{\tilde{l}}^{\alpha t_k} d l' \frac{1+\frac{l'}{2\pi R_c}}{\Gamma(l')}. \label{tksolve}
\end{align}
Here the integration variable $l'$ can be interpreted as the instantaneous length of the loop emitting gravitational wave at time $t'$, where $t_k < t' < \tilde{t}$. We further assume that at time $\tilde{t}$, the newly created loops are larger than the pre-existing loops emitting gravitational wave at time $\tilde{t}$
\begin{align}
    \alpha \tilde{t} > \tilde{l}.
\end{align}

The energy density of the stochastic gravitational wave spectrum at time $t$ is given by
\begin{align}
    \frac{d\rho_{\rm GW}(t)}{df} &= \int_{t_{\rm sc}}^{t} d\tilde{t} \left(\frac{a(\tilde{t})}{a(t)}\right)^4 \int dl \frac{dn(l, \tilde{t})}{dl} \frac{dP(l, \tilde{t})}{d\tilde{f}} \frac{d\tilde{f}}{df}. \label{drhodf}
\end{align}
Here $\tilde{t}$ is the emission time of the gravitational wave, and $t_{\rm sc} \sim 10^{-22}$ s is the time when the network reaches a scaling regime where the energy density of the hybrid network becomes constant in the expanding Universe. 
The first term of the rightmost integrand of Eq.~\eqref{drhodf} can be decomposed as 
\begin{align}
    \frac{dn(l, \tilde{t})}{dl} = \frac{dn}{dt_k} \frac{dt_k}{dl}. \label{dndl}
\end{align}
The second term in Eq.~\eqref{dndl} is determined by differentiating Eq.~\eqref{tksolve} with respect to $t_k$,
\begin{align}
    \frac{dt_k}{dl} = \frac{1+\frac{l}{2\pi R_c}}{\Gamma(l) G\mu} \left[ 1 + \frac{\alpha \left( 1 + \frac{\alpha t_k}{2\pi R_c}\right)}{\Gamma(\alpha t_k)G \mu} \right]^{-1},
\end{align}
while the first term represents the loop number density production rate calculated from the one-scale model and calibrated from simulations \cite{Cui:2018rwi, Gouttenoire:2019kij, Sousa:2013aaa}
\begin{align}
    \frac{dn}{dt_k} &= \left[ \frac{\mathcal{F} C_{\rm eff}(t_k)}{\alpha t_k^4} \left(\frac{a(t_k)}{a(\tilde{t})}\right)^3 \right] \theta{(t_\star - t_k)}.
\end{align}
We have assumed that at every Hubble time, roughly one loop of size $\alpha t_k$ breaks off from the infinite wall-string network, and then it redshifts as $a^{-3}$, $a(t)$ being the scale factor. $\mathcal{F} \approx 0.1$ is the fraction of energy that is transferred by the string network into loops \cite{Blanco-Pillado:2013qja}. $C_{\rm eff}$ is the loop formation efficiency which equals $5.7$ during the radiation-dominated era, and $0.5$ during the matter-dominated era \cite{Cui:2017ufi, Blasi:2020wpy}. The Heaviside $\theta$ function $\theta(t_\star - t_k)$ ensures that we only consider the contribution from loops which were created before the network collapses at $t_\star$.

The second term in the rightmost integrand of Eq.~\eqref{drhodf} can be expressed as
\begin{align}
    \frac{dP(l, \tilde{t})}{d\tilde{f}} &= \Gamma(l) G\mu^2l\ g(\tilde{f}l).
\end{align}
It is useful to decompose the gravitational wave into Fourier modes $\tilde{f}_n = \xi n/\tilde{l}$, where $\xi \equiv l/T$ varies between $2$ in the pure string limit ($\tilde{l} \ll 2\pi R_c$) and $\pi$ in the pure wall limit $\tilde{l} \gg 2\pi R_c$ \cite{Dunsky:2021tih}.. The normalized power spectrum for a discrete spectrum is then given by
\begin{align}
    g(x) &= \sum_n \mathcal{P}_n\ \delta(x - \xi n). \label{gx}
\end{align}
Here $\mathcal{P}_n = n^{-q}/\zeta(q)$ is the fractional power radiated by the $n$th mode of an oscillating string loop with cusp, and $\zeta(x) = \sum_{m=1}^{\infty} m^{-x}$ is the Riemann zeta function. The power spectral index is found to be $q = 4/3$ for string loops containing cusps \cite{Auclair:2019wcv, Vachaspati:1984gt}. 

Finally, accounting for redshifting, the emission frequency $\tilde{f}$ at time $\tilde{t}$ is related to the observed frequency $f$ at time $t$ by
\begin{align}
    f = \tilde{f}\ \frac{a(\tilde{t})}{a(t)}, \label{fredshift}
\end{align}
where the scale factor $a(t)$ is defined as $a(t_0) \equiv 1$ for present time $t_0$. From Eq.~\eqref{fredshift} the third term of the rightmost integrand in Eq.~\eqref{drhodf} can be easily obtained.

Putting everything together, the stochastic gravitational wave spectrum observed today is given by summing over all Fourier modes
\begin{align}
    \Omega_{\rm GW} (f) &\equiv \frac{f}{\rho_c} \frac{d\rho_{\rm GW}}{df} \\
    &= \sum_n \frac{G\mu^2}{\rho_c} \int_{t_{\rm sc}}^{t_0} d\tilde{t} \left( \frac{a(\tilde{t})}{a(t_0)}\right)^5 \left[ \frac{\mathcal{F} C_{\rm eff}(t_k)}{\alpha t_k^4} \left(\frac{a(t_k)}{a(\tilde{t})}\right)^3 \right] \mathcal{P}_n \frac{\xi n}{f} \left[ 1+ \frac{1}{2\pi R_c } \frac{\xi n}{f}\frac{a(\tilde{t})}{a(t_0)} \right] \nonumber \\
    &\times \frac{\Gamma(\alpha t_k) \theta{(t_\star - t_k)}}{\Gamma(\alpha t_k) G\mu + \alpha \left(1+\frac{\alpha t_k}{2\pi R_c}\right)}. \label{GWformula}
\end{align}
Here $\rho_c = 3H_0^2/(8\pi G)$ is the critical energy density today. The integral with respect to $l$ in Eq.~\eqref{drhodf} is performed utilizing the $\delta$ function in Eq.~\eqref{gx} to replace $l = \frac{\xi n}{f} \frac{a(\tilde{t})}{a(t_0)}$ for each $n$-mode.

In Eq.~\eqref{GWformula}, we are integrating over the gravitational wave modes emitted by the hybrid network from an early time when the network goes to a scaling regime, $t_{\rm sc}$, to today $t_0$. Here the integration variable $\tilde{t}$ represents the time when a loop is radiating gravitational wave, and $t_k$ is the time when this particular loop was created, which can be obtained by solving Eq.~\eqref{tksolve}.   

If no domain walls were created (when the $U(1)_R$ symmetry is broken to nothing), we could take the $R_c \rightarrow \infty$ limit, and Eq.~\eqref{GWformula} would yield the gravitational wave spectrum generated from the decay of a pure string network. In this case, the Heaviside theta function would be replaced by $\theta(t_k - t_{\rm sc})$ to imply that we only consider the loops created after $t_{\rm sc}$ when the network reaches a scaling regime, and loop creation does not stop at any characteristic time similar to $t_{\star}$. This distinction has important consequences for the infrared tail of the GW spectrum, as will be discussed later.

We have developed a public code package \textsc{CosmicStringGW}, available on GitHub at {\href{https://github.com/moinulrahat/CosmicStringGW}{\texttt{\faGithubSquare/moinulrahat/CosmicStringGW}}}, for calculating the gravitational wave spectrum from pure and hybrid defect networks of cosmic strings.

\section{Gravitational wave signature\label{sec:GW}} 
\begin{figure}[t!]
    \centering
    \includegraphics[width=1.1\textwidth]{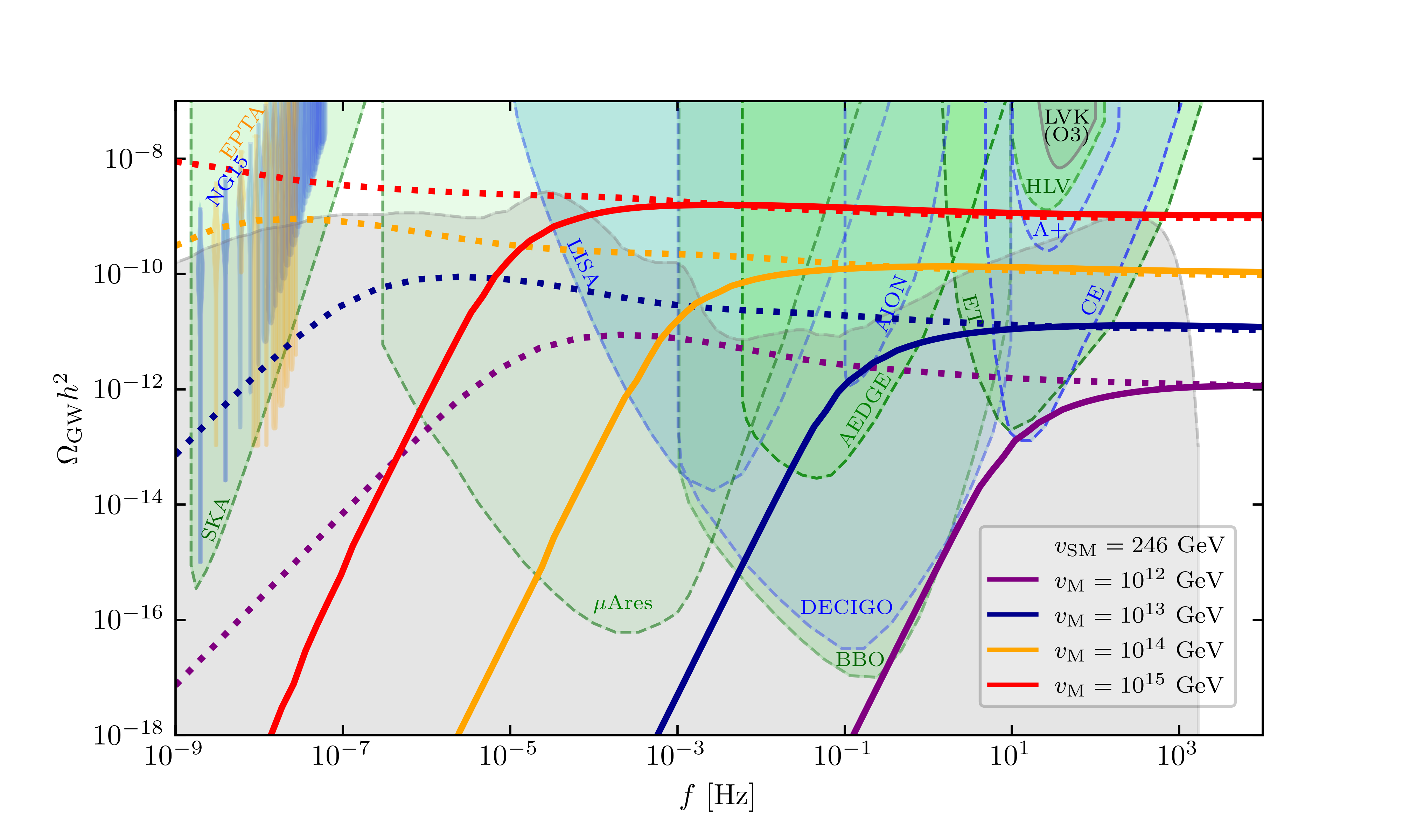}
    \caption{GW spectrum from domain walls bounded by cosmic strings (solid lines). Dotted lines represent the corresponding pure string induced GW spectrum in the absence of an intermediate $Z_2$ breaking. Sensitivity of various upcoming interferometers are shown with colored dashed lines. Solid gray region on top right corner shows upper bound from LVK third observing run. Gray region bounded by dashed line on bottom part of the plot represent the sum of various astrophysical foregrounds. We show known foregrounds up to $1.7 \times 10^3$ Hz; however, there may be other foregrounds in higher frequencies.  See text for details.}
    \label{fig:GW2hdm1}
\end{figure}

We expect the $U(1)_R$ symmetry to be spontaneously broken at a high scale above $10^{10}$ GeV, inspired by a high scale of right-handed neutrino masses, and the $Z_2$ symmetry to be broken at the electroweak scale. Hence we focus on keeping the $Z_2$ breaking scale fixed at $v_{\rm SM} = 246$ GeV, while varying the $U(1)_R$ breaking scale. The resulting gravitational waves from the string-wall network, calculated from evaluating Eq.~\eqref{GWformula} up to $k = 10^5$ terms, is shown in Fig.~\ref{fig:GW2hdm1} with solid lines. In the same plot we also show the spectrum corresponding to the pure string case (achieved by taking the $R_c \to \infty$ limit in Eq.~\eqref{GWformula}), if the $U(1)_R$ symmetry were completely broken at the high scale. Projected sensitivity (SKA \cite{Janssen:2014dka}, $\mu$-Ares \cite{Sesana:2019vho}, LISA \cite{LISA:2017pwj}, DECIGO \cite{Kudoh:2005as, Kawamura:2020pcg}, BBO \cite{Harry:2006fi}, AEDGE \cite{AEDGE:2019nxb}, AION \cite{Badurina:2019hst}, CE \cite{LIGOScientific:2016wof}, ET \cite{Hild:2008ng} and future upgrades of LVK), upper bounds (LVK \cite{LIGOScientific:2022sts, KAGRA:2021kbb, Jiang:2022uxp}) from various interferometers and recent results from pulsar timing arrays NANOGrav \cite{NANOGrav:2023gor, NANOGrav:2023hvm} and EPTA \cite{EPTA:2023sfo, EPTA:2023fyk} are shown for comparison. Astrophysical foregrounds for the relevant frequency range are also shown, and will be discussed later in this section. Signal-to-noise ratio (SNR) at some of the detectors is tabulated in Appendix~\ref{app:snr}.

We notice the following key characteristics of the gravitational wave spectrum generated from the hybrid network of domain walls bounded by cosmic strings. 
\begin{itemize}
    \item Compared to the pure string case (dotted line), the signals from the hybrid network (solid line) have an infrared (IR) tail demonstrating a departure from the flat spectrum occurring at larger frequencies. This departure is caused by the collapse of the hybrid network after creation of the domain walls that drive the boundary strings to ultra-relativistic dynamics. The frequency where the IR tail starts is determined by the collapse time $t_{\rm DW}$, which in our model depends on $R_c \sim v_{\rm{M}}^3 / v_{\rm SM}^2$ (since $R_c > t_{\rm DW}$ for $v_{\rm {M}} \gtrsim 10^{10}$ GeV). Higher $R_c$, represented by higher $U(1)_R$ breaking scale for a fixed $Z_2$ breaking scale considered here, implies that the collapse happens later, allowing the flat part of the signal to continue to lower frequencies (see Fig.~\ref{fig:GW2hdm1}).
    
    \item The infrared spectral slope of the hybrid-defect signal is steeper compared to the pure string case, featuring a $f^3$ power-law typical of instantly decaying sources arising from causality arguments \cite{Caprini:2009fx, Cai:2019cdl, Brzeminski:2022haa}. This phenomenon can be understood from the dynamics of the string bounded walls discussed in Sec.~\ref{sec:hybrid}. Similar to the pure string case, the dynamics of the hybrid network at early time (corresponding to higher frequencies) is dominated by the non-relativistic string loops. This represents the situation after the $U(1)_R$ breaking and before the $Z_2$ breaking. After the $Z_2$ breaking, domain walls appear and get bounded by the pre-existing string loops. The domain wall formation time $t_{\rm DW}$ is controlled by the $Z_2$ breaking scale $v_{\rm SM}$. Walls do not immediately affect the ongoing string-dominated dynamics if $t_{\rm DW} < R_c$, where $R_c \sim v_{\rm M}^2 / v_{\rm SM}^3$ is controlled by the separation of the two breaking scales. However, at $t_\star \equiv \text{max}(R_c, t_{\rm DW}) \sim R_c$ for $v_M \gtrsim 10^{10}$ GeV, two important phenomena occur that dictates the fate of the GW spectrum at infrared scales. First, new loop formation ceases, hence only the pre-existing loops can decay and contribute to the GW spectrum. Second, the pre-existing loops become ultra-relativistic, leading to a higher chopping rate and near-instantaneous decay. Overall, the situation mimics the case of a source that decays very quickly, and the resulting spectrum demonstrates a $f^3$ slope, as pointed out in Refs.~\cite{Caprini:2009fx, Cai:2019cdl, Brzeminski:2022haa}. Note that the condition of no loop creation after $t_{\star}$ has been explicitly put into the Eq.~\eqref{GWformula} with the Heaviside theta function, marking a striking difference with the pure string case, where loop formation continues resulting in a milder slope of the spectrum at IR frequencies. The steep $f^{3}$ slope of the IR tail in the present case can be distinguished from the pure string case in planned interferometer sensitivities in the $\mu$Hz to Hz frequencies.

    \item The IR tail appearing at relatively higher frequencies compared to the pure string case implies that non-observation at pulsar timing arrays (PTA) does not rule out signals from the hybrid defect for $v_{\rm M} \gtrsim 10^{14}$ GeV. The upper bound $v_{\rm M} \lesssim 10^{15.5}$ GeV from LVK applies to gravitational waves from both pure and hybrid cosmic string defects, as both signals are identical at LVK frequencies.
    
    \item While the pure string signal probes a single symmetry breaking scale \cite{Dror:2019syi, Dasgupta:2022isg, Fu:2023nrn, DiBari:2023mwu, King:2023cgv}, signals from hybrid defects like domain walls bounded by cosmic strings can simultaneously probe two symmetry breaking scales ($U(1)_R$ and $Z_2$). 
    This offers a unique opportunity to test BSM scenarios where such scales are separated by orders of magnitude.

    \item Given that the $Z_2$ breaking scale in the current model is fixed at the electroweak scale, the model predicts a signal with spectral slope $f^{3}$ in the $\mu$Hz to Hz range for $U(1)_R$ breaking scale $10^{12} - 10^{15}$ GeV. Together with a flat spectrum at frequencies higher than the IR tail, this is a unique signature of this model.
\end{itemize}

\subsection{The astrophysical foreground} 

One may expect astrophysical foreground from various sources which could contribute to stochastic GW background. To name a few, the already observed binary black hole (BH-BH)  \cite{TheLIGOScientific:2016qqj, Abbott:2016nmj, TheLIGOScientific:2016pea, Song:2024pnk} and 
binary neutron star (NS-NS) by LIGO/VIRGO \cite{TheLIGOScientific:2017qsa}.
In order to extract the signal and distinguish between the SGWB sourced walls bounded by strings as we predicted in our analysis and those from the one generated by the astrophysical foreground, one should be able to subtract the astrophysical signals expected with sensitivities of BBO and ET or CE windows \cite{Cutler:2005qq,Regimbau:2016ike}.
Besides, there could be the binary white dwarf galactic and extra-galactic astrophysical foreground also present in LISA as the dominant component as shown in Refs. \cite{Farmer:2003pa, Rosado:2011kv, Moore:2014lga} and should be subtracted \cite{Kosenko:1998mv,Adams:2010vc, Adams:2013qma}. In our analysis we assume that
such subtractions would be done in the future. Together with the crucial fact that the GW spectrum generated by the astrophysical foreground increases with frequency as $f^{2/3}$ (as shown in Ref. \cite{Zhu:2012xw}), the spectral shapes of which is completely different from the GW spectrum from walls bounded by strings as we showed, as suggested by Eq.~\eqref{GWformula} and Fig.~\ref{fig:GW2hdm1}, we hope to zero in on the GW signals from topological defects precisely. We show the sum of various astrophysical foregrounds in the gray shaded region bounded by a dashed line in the lower part of Fig.~\ref{fig:GW2hdm1}. Notice that the known foregrounds are considered here up to frequencies $1.7 \times 10^3$ Hz, but new sources may contribute to foregrounds at higher frequencies. However, a detailed analysis of such subtraction and estimation is beyond the scope of the present paper and we leave it for future publication. 

In summary, the hybrid network generated  from the sequential breaking of the $U(1)_R$ symmetry via the intermediate $Z_2$ symmetry results in a unique GW signal distinguishable from other new physics signals in the observable frequency range, and can be probed in upcoming GW interferometers.

\section{Conclusion}\label{sec:conclusion}

We have investigated the emergence of a hybrid topological defect, namely ``domain walls bounded by cosmic strings'', and the unique gravitational wave signal that emerges from it in the context of the type-I two-Higgs-doublet model (2HDM) extended with a gauged $U(1)_R$ symmetry. The phenomenological motivation for the $U(1)_R$ extension stems from the necessity to incorporate neutrino mass and mixing via the seesaw mechanism. The model simultaneously cancels the chiral anomaly and eliminates the flavour-changing neutral currents. An interesting outcome of the $U(1)_R$ charge assignment, motivated by spontaneous generation of the right-handed neutrino Majorana masses, is that $U(1)_R$ is broken to an intermediate $Z_2$ symmetry, which is further broken around the electroweak scale by the Higgs doublets. Such a chain of symmetry breaking triggers a hybrid topological defect of ``domain walls bounded by cosmic strings'', created by the domain walls of $Z_2$ breaking being encircled by pre-existing cosmic string loops originated from the earlier $U(1)_R$ breaking.

We have studied the cosmological footprint of this model in terms of the decay of the hybrid network of walls bounded by strings into gravitational waves. The defect network evolves by losing energy via gravitation wave emission, where the observable frequency of the signal is inversely proportional to the time of evolution.  The decay of the cosmic string loops created after $U(1)_R$ breaking yields a flat spectrum of gravitational waves at high frequencies. The additional tension exerted by the bounded domain walls arising after $Z_2$ breaking causes an accelerated decay of the loops, resulting in a characteristic $f^{3}$ slope of the signal in lower frequencies. The pivot frequency where this spectral change occurs is determined by the ratio of the two symmetry breaking scales, while the high frequency flat amplitude is representative of the higher scale. In the present model, the $Z_2$ is broken around the electroweak scale. This leads to a prediction for the $f^{3}$ behaviour to be observed in the microHz to Hz band with a flat spectrum in the ultraviolet, when the $U(1)_R$ breaking scale is in between $10^{12}$ GeV to $10^{15}$ GeV. In this sense, our study provides a way to test the model in upcoming gravitational wave interferometers such as LISA and ET, complementing the collider results.

\section*{Acknowledgement}
We thank Stefan Antusch, Yongcheng Wu and Urjit Yajnik for helpful discussions. MHR acknowledges financial support from the Generalitat Valenciana project CIPROM/2021/054 and Spanish AEI-MICINN PID2020-113334GB-I00/AEI/10.13039/501100011033.
SFK thanks CERN for hospitality and acknowledges the STFC Consolidated Grant ST/L000296/1 and the European Union's Horizon 2020 Research and Innovation programme under Marie Sk\l{}odowska-Curie grant agreement HIDDeN European ITN project (H2020-MSCA-ITN-2019//860881-HIDDeN). 

\appendix
\section{The scalar sector of 2HDM\label{app:scalars}}
The minimisation conditions for the Higgs potential are given by 
\begin{subequations}
\begin{align}
    \frac{\partial V}{\partial v_1} &= \tilde{m}_{11}^2 v_1 - \tilde{m}_{12}^2 v_2 + \frac{\lambda_1}{2} v_1^3 + \frac{\lambda_{34}}{2} v_1 v_2^2 =0 \,,\label{eq:minicon1}\\
    \frac{\partial V}{\partial v_2} &= \tilde{m}_{22}^2 v_2 - \tilde{m}_{12}^2 v_1 + \frac{\lambda_2}{2} v_2^3 + \frac{\lambda_{34}}{2} v_1^2 v_2 =0 \,,\label{eq:minicon2}
\end{align}
\end{subequations}
where $\lambda_{34} = \lambda_3 + \lambda_4$. 
After the spontaneous symmetry breaking, there are five physical scalar fields: two neutral scalars ($H$ and $h$), a charged scalar ($H^\pm$), and one pseudoscalar ($A$). The mass spectrum of the scalar particles are
\begin{align}
    m_\pm^2 &= \tilde{m}_{12}^2 (\tan\beta + \cot\beta) - \lambda_4 v_{\rm SM}^2 \,,\label{eq:mpm}\\
    m_A^2 &= \tilde{m}_{12}^2 (\tan\beta + \cot\beta) \,, \label{eq:mA}\\
    \begin{pmatrix}
        m_H^2 & 0 \\
        0 & m_h^2
    \end{pmatrix}
    &=
    \begin{pmatrix}
        c_\alpha & s_\alpha \\
        -s_\alpha & c_\alpha
    \end{pmatrix}
    \begin{pmatrix}
        \tilde{m}_{12}^2 \tan\beta + \lambda_1 v_1^2 & 
        -\tilde{m}_{12}^2 + \lambda_{34} v_1 v_2 \\
        -\tilde{m}_{12}^2 + \lambda_{34} v_1 v_2 &
        \tilde{m}_{12}^2 \cot\beta + \lambda_2 v_2^2
    \end{pmatrix}
    \begin{pmatrix}
        c_\alpha & -s_\alpha \\
        s_\alpha & c_\alpha
    \end{pmatrix}\,,\label{eq:mhiggs}
\end{align}
where $c_\alpha=\cos\alpha$ and $s_\alpha=\sin\alpha$.

In total, there are 7 model parameters and 7 physical parameters in the Higgs sector, as listed in Table~\ref{tab:parameters}. 
\begin{table}[t!]
\centering
\renewcommand\arraystretch{1.1}
\begin{tabularx}{0.58\textwidth}{ c c }
\toprule
Model parameters & Physical parameters \\ \midrule
$\tilde{m}_{11,22,12},\, \lambda_{1,2,3,4}$ & $m_\pm,\, m_A,\, m_H,\, m_h,\, v_{\rm SM}\,, \beta,\, \alpha$\\
\bottomrule
\end{tabularx}
\caption{\label{tab:parameters} Parameters in the Higgs sector.}
\end{table}
The physical parameters and the model parameters are related by 
\begin{align}
    \tilde{m}_{11}^2 &= \frac{1}{2} \sec\beta 
    (- m_H^2 \cos\alpha\cos(\alpha-\beta) - m_h^2 \sin\alpha \sin(\alpha-\beta) + m_A^2 \sin\beta \sin2\beta) \,,\\
    \tilde{m}_{22}^2 &= \frac{1}{2} \csc\beta 
    (- m_H^2 \sin\alpha \cos(\alpha-\beta) + m_h^2 \cos\alpha \sin(\alpha-\beta) + m_A^2 \cos\beta \sin2\beta) \,,\\
    \tilde{m}_{12}^2 &= \frac{m_A^2}{\tan\beta+\cot\beta}\,,\\
    \lambda_{1} &= \frac{1}{v_{\rm SM}^2 \cos^2\beta}\left(m_H^2 \cos^2\alpha + m_h^2 \sin^2\alpha - m_A^2 \sin^2\beta \right) \,,\\
    \lambda_{2} &= \frac{1}{v_{\rm SM}^2 \sin^2\beta}\left(m_H^2 \sin^2\alpha + m_h^2 \cos^2\alpha - m_A^2 \cos^2\beta \right)\,,\\
    \lambda_{3} &= \frac{1}{v_{\rm SM}^2\sin2\beta }(m_H^2-m_h^2)\sin2\alpha + \frac{m_\pm^2}{v_{\rm SM}^2}\,,\\
    \lambda_{4} &= \frac{m_A^2-m_\pm^2}{v_{\rm SM}^2}\,.
\end{align}
In the alignment limit, these relations between model and physical parameters are reduced to 
\begin{align}
    &\tilde{m}_{11}^2 = \frac{1}{2}  
    (- m_h^2 + 2 m_A^2 \sin^2\beta) \,,&\quad
    &\tilde{m}_{22}^2 = \frac{1}{2} 
    (- m_h^2 + 2 m_A^2 \cos^2\beta) \,,&\quad
    \tilde{m}_{12}^2 = \frac{m_A^2}{\tan\beta+\cot\beta}\,,& \nonumber\\
    &\lambda_{1} = \frac{(m_H^2 - m_A^2) \tan^2\beta + m_h^2}{v_{\rm SM}^2 } \,,&\quad
    &\lambda_{2} = \frac{(m_H^2 - m_A^2) \cot^2\beta + m_h^2}{v_{\rm SM}^2 } \,,&&&\nonumber\\
    &\lambda_{3} = \frac{m_\pm^2-m_H^2+m_h^2}{v_{\rm SM}^2}\,,&\quad
    &\lambda_{4} = \frac{m_A^2-m_\pm^2}{v_{\rm SM}^2}\,.&&& 
    \label{eq:SMlimit}
\end{align}
The stability of the vacuum can be ensured by $\lambda_{1,2}>0$, $\lambda_3>-\sqrt{\lambda_1 \lambda_2}$ and $\lambda_3 + \lambda_4>-\sqrt{\lambda_1 \lambda_2}$.

\section{Equations of motion for domain wall in the Higgs basis \label{app:basis}}
In the 2HDM, there is a basis of the Higgs doublets under which only one of the Higgs doublets obtains a VEV, which is known as the Higgs basis. 
The Higgs basis can be obtained by the following rotation of the fields:  
\begin{eqnarray}
    \Phi_1\rightarrow S_1 \cos\beta - S_2 \sin\beta\,,\quad
    \Phi_2\rightarrow S_1 \sin\beta + S_2 \cos\beta\,.
\end{eqnarray}
In the Higgs basis, the scalar potential Eq.~\eqref{eq:potential} would be minimised at $(s_1,s_2)=(v_{\rm SM},0)$, where $s_i\equiv \sqrt2\langle S_i \rangle$.

Under the alignment limit, the path of the field value for the domain wall solution happens to be a straight line connecting the two $Z_2$ symmetric vacua. Along the $S_1$ direction ($s_2=0$), the gradient of the potential reads 
\begin{eqnarray}
    \left.\frac{\partial V}{\partial s_1}\right|_{s_2=0} = \frac{m_h^2}{2 v_{\rm SM}^2} s_1 (s_1^2 -  v_{\rm SM}^2)\,,\quad
    \left.\frac{\partial V}{\partial s_2}\right|_{s_2=0} = 0\,,
\end{eqnarray}
where $m_h$ is the mass of $h$.
Therefore the equation of motion for $s_2$, which is $d^2s_2/dz^2=\partial V/\partial s_2$, is automatically satisfied when $s_2=0$. 
In this case, the domain wall solution can be obtained by solving the equation of motion for $s_1$.

However, in the more general case without the alignment limit, it is not necessary that only one Higgs boson gets VEV everywhere in the domain wall solution for 2HDM. 
In that case, the gradient of the potential along the $S_1$ direction ($s_2=0$) is 
\begin{eqnarray}
    \left.\frac{\partial V}{\partial s_1}\right|_{s_2=0} &=& \frac{m_h^2+m_H^2+(m_H^2-m_h^2)\cos 2(\alpha-\beta)}{2 v_{\rm SM}^2} s_1 (s_1^2 - v_{\rm SM}^2)\,,\\
    \left.\frac{\partial V}{\partial s_2}\right|_{s_2=0} &=& \frac{(m_H^2-m_h^2)\sin 2(\alpha-\beta)}{2 v_{\rm SM}^2} s_1 (s_1^2 - v_{\rm SM}^2)\,.
\end{eqnarray}
The equation of motion for $s_2$ is no longer trivial when $s_2=0$.
In order to get the correct domain wall solution, the path has to depart from the $S_1$ direction. 
Although only $S_1$ gets a VEV at the minima, the value of $S_2$ is non-zero inside the domain wall. 

\section{Signal-to-noise ratio} \label{app:snr}
We calculate the signal-to-noise ratio (SNR) of the gravitational wave signal from the hybrid topological defect at various interferometers as follows \cite{Caprini:2015zlo}.
\begin{align}
    \text{SNR} &= \sqrt{\tau \int_{f_{\rm min}}^{f_{\rm max}} df 
    \left[ \frac{\Omega_{\rm GW}h^2(f)}{\Omega_{\rm Sens}h^2}\right]^2},
\end{align}
where $\Omega_{\rm Sens}h^2$ denotes the sensitivity of the interferometers, $\Omega_{\rm GW}h^2$ is the GW signal, and $\tau$ is the duration of running in years. We take $\tau = 4$ for all interferometers. In this definition, $\text{SNR} > 10$ is taken to be the threshold for detection in a given interferometer \cite{Caprini:2015zlo}. The resulting SNR is shown in Table~\ref{tab:SNR}.

We have considered the power law integrated sensitivity implying that we denote a signal to be detectable when its amplitude surpasses the power-law sensitivity curve for a given signal to-noise ratio (SNR), which have been estimated following that the GW signal assumes a power law \cite{Thrane:2013oya}. It is usually assumed that among the astrophysical foregrounds present, supermassive black hole binaries can be resolved while the compact binaries cannot be, see Ref.~\cite{Breitbach:2018ddu} for further details associated to this. However, we remark that the expression for SNR used is an approximated one which is consistent within the limit of a large detector noise \cite{Gouttenoire:2019kij,Gouttenoire:2021jhk,DelleRose:2019pgi,Conaci:2024tlc}. The generic expression for SNR can be found in Refs.~\cite{Allen:1997ad,Kudoh:2005as,Brzeminski:2022haa} and involves more detailed analysis which is beyond the scope of the current work.

\begin{table}[htbp]
\centering
\renewcommand\arraystretch{1.1}
\begin{tabularx}{0.97\textwidth}{c c c c c}
\toprule
{Interferometer} & {${v_{\rm{M}}} = 10^{12}$} GeV & $v_{\rm{M}} = 10^{13}$ GeV & $v_{\rm{M}} = 10^{14}$ GeV & $v_{\rm{M}} = 10^{15}$ GeV\\
\midrule
$\mu$ARES & -- & -- & $6.9\times 10^3$ & $1.15 \times 10^6$ \\
LISA & -- & 0.01 & 343.51 & $10^4$ \\
BBO & 3.46 & $3.05\times 10^5$ & $1.28\times 10^7$ & $1.36\times 10^8$ \\
DECIGO & 2.43 & $7.13\times 10^4$ & $3.51\times 10^6$ & $3.85\times 10^7$ \\
AION & -- & 0.97 & 57.51 & 630.99 \\
AEDGE & -- & 15.19 & $2.08\times 10^3$ & $2.61 \times 10^4$ \\
ET & 7.21 & 379.58 & $4.33\times 10^3$ & $3.78\times 10^4$ \\
CE & 12.79 & 654.21 & $7.31\times 10^3$ & $6.38\times 10^4$ \\
A+ & 0.01 & 0.24 & 2.58 & 22.55 \\
HLV & -- & 0.07 & 0.71 & 6.25 \\
\bottomrule
\end{tabularx}
\caption{Signal-to-noise ratio for $v_{\rm SM} = 246$ GeV. Missing values denoted by `--' imply our inability to resolve the signal from astrophysical foregrounds.}
\label{tab:SNR}
\end{table}

\newpage
\bibliography{Ref}
\bibliographystyle{JHEP}

\end{document}